\documentclass{ceab}  %% loading ceab.cls creates ceab style
\usepackage{epsfig}
\usepackage{xfrac}
\usepackage{amssymb}

\def\gore{W. Curdt and B. Fleck}

\begin{document}

\title{Solar and Galactic Cosmic Rays Observed by SOHO}

\author{\normalsize W. CURDT$^1$, B. FLECK$^2$ \vspace{2mm} \\
        $^1$\it Max Planck Institute for Solar System Research \\
            \it Justus-von-Liebig-Weg 3, 37077 G\"ottingen, Germany\\
        $^2$\it ESA Science Operations Department, c/o NASA/GSFC,\\
            \it Greenbelt, MD, USA}
\maketitle

\begin{abstract}
Both the Cosmic Ray Flux (CRF) and Solar Energetic Particles (SEPs) have left an
imprint on SOHO technical systems. While the solar array efficiency degraded
irreversibly down to $\approx$77\% of its original level over roughly 1\,\sfrac{1}{2}
solar cycles, Single Event Upsets (SEUs) in the solid state recorder (SSR) have
been reversed by the memory protection mechanism. We compare the daily CRF
observed by the Oulu station with the daily SOHO SEU rate and with the degradation
curve of the solar arrays. The Oulu CRF and the SOHO SSR SEU rate are both modulated
by the solar cycle and are highly correlated, except for sharp spikes in the SEU rate,
caused by isolated SEP events, which also show up as discontinuities in the otherwise
slowly decreasing solar ray efficiency. This allows to discriminate between effects
with solar and non-solar origin and to compare the relative strength of both.
We find that during solar cycle 23  (1996 Apr 1 -- 2008 Aug 31) only 6\% of the total
number of SSR SEUs were caused by SEPs; the remaining 94\% were due to galactic cosmic rays.
During the maximum period of cycle 23 (2000 Jan 1 -- 2003 Dec 31), the SEP contribution
increased to 22\%, and during 2001, the year with the highest SEP rate, to 30\%.
About 40\% of the total solar array degradation during the 17 years from Jan 1996
through Feb 2013 can be attributed to proton events, i.e. the effect of a series
of short-lived, violent SEP events is comparable to the cycle-integrated damage by cosmic rays.
\end{abstract}


\section{Introduction}
\large
SOHO was launched on 1995 Dec 2, and inserted in a halo orbit around the
first Lagrangian point, L$_1$, in Feb 1996. With the exception of a longer interruption
in 1998 and a shorter one in early 1999 it has been operational ever since.
Not protected by geomagnetic, geocoronal or atmospheric shielding, the instruments ---
but also the technical systems of SOHO --- have been
exposed to space environment. Here we focus on effects caused by the cosmic
ray flux (CRF) and those stemming from Solar Energetic Particles (SEPs) during
the entire solar cycle 23 and half of cycle 24. Both CRF and SEPs leave a signature in the
memory protection system of SOHO's Solid State Recorder (SSR). It turned out
that --- beside its technical benefit --- this device is also of scientific
use, since it can be used as a reliable monitor to measure energetic particles.
This was already noticed by McIntosh et al. (2013). In their
work on the phase relationship between magnetic activity of the two hemispheres
and its effect on the solar cycle behaviour (an entirely different scientific question than
that discussed here) they compared the CRF measured by the Oulu Neutron Monitor (ONM)
at the Sodankula Geophysical Observatory with SSR SEU records obtained by SOHO.
Here, we extend that work and refer also to those data points that appear as outliers in their Fig.~2.
Furthermore, we present interesting details about the correlation between both data
sets. Finally, we present the imprint of SEPs and CRF on the solar array power system.
The motivation for this work was to establish a quantitative relationship
between cosmic ray effects and solar effects in the energy range responsible for
the observed phenomena.

\section{Solid State Recorder Single Event Upsets}

SOHO is equipped with a solid state memory that buffers telemetry during
non-contact hours with a full capacity of 2 G-bit. The memory chips are
Texas Instruments SMJ44100 process S2.1 4Mx1 DRAMs.
The SSR is protected against latch-ups and data corruption caused by Single-Event-Upsets (SEUs).
Such a protection --- required in the harsh environment of outer
space --- is provided by a logic that is based on a Hamming code algorithm and
detects, corrects and logs such bitflips at a rate of 15 seconds.
This data protection employs a scrubbing function that continuously `cleans'
the memory by reading the data stored in memory and correcting it if needed.
It takes 13~$\mu$s to check one memory word,
therefore, each 16-bit word is read every 29 minutes when the full capacity, 2 G-bit,
(128 $\times$ 1024 $\times$ 1024 words of 16 bits) is used.
Each error-corrected SEU also increments an SEU counter.
In a worst case of 1 single-bit error in each memory word, the maximum rate reported
would be very high, namely 4.6~10$^6$ counts per minute (cpm). Under normal conditions
the observed rate is much lower (0.5 to 1.3 cpm). However, during violent solar storms,
the 8-bit counter sometimes was overrun within 15~s and the actual rate had to be
determined manually. Each RAM chip contains only a single bit of a data word in memory.
Hence only in case of errors at the same location in two different chips belonging
to the same 16-bit word one would get a double error (note that one was reported on 1997
Nov 13). In a way a double error is a measurement of a higher rate in radiation.
A description of this device and the statistics of SOHO-wide SEU events is given by
Harboe-S{\o}rensen et al. (2002) from an engineering perspective.

\subsection{Observations}

In Fig.~1 we show the Oulu CRF as measured by the ONM (black) and the data of
the SOHO SSR (red), given in units of counts per minute (cpm). Both axes have been
scaled using a linear regression between both data sets (see also Fig.~2)
and clipped in a way that the graphs can be easily compared.

\begin{figure} [htb] % Fig 1 Oulu vs SEU
   \centering
   \includegraphics[width=11cm]{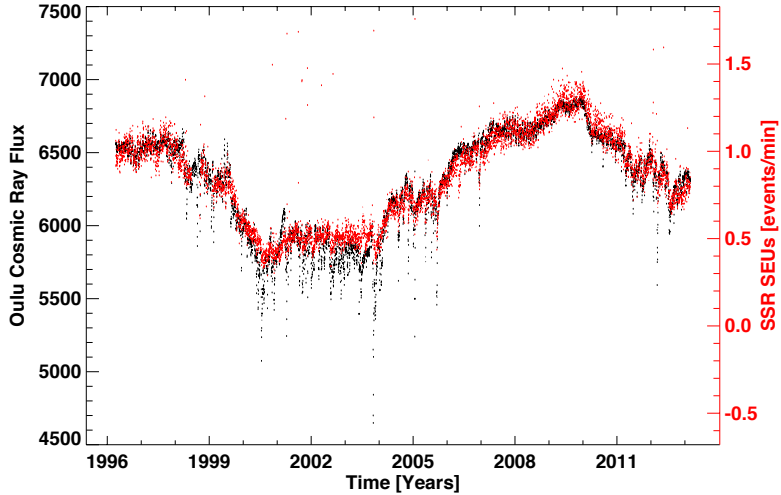}
    \caption{Daily average Single Event Upset rate (red) and Oulu Cosmic Ray Flux
      (black) from Apr 1996 to Feb 2013. The SEU axis is clipped such
      that the very good overall agreement of both graphs becomes obvious.
      }
\label{SEU}
\end{figure}

The overall good correlation over 17 years (1\,\sfrac{1}{2} solar cycles) is obvious,
demonstrating the CRF nature of the SEU background.
There are, however, remarkable differences.
The magnitude of the solar cycle variation in SOHO SSR SEU data is significantly
larger than that of the ONM CRF. It increases from about
0.5 cpm to over 1.3 cpm during the recent solar minimum (factor 2.5).
The Oulu CRF, on the other hand, varies from 5900 to 6800 daily events
from solar maximum to solar minimum (this is about 15\%). We attribute this difference
to the different energy cross section of both measurement systems.
The ONM measures solely the very energetic CRF particles which are
less modulated by the heliospheric magnetic field variations over the solar
cycle. We note that in both data sets the second minimum in 2009 was much more
pronounced than the 1996 minimum. This anomalously high CRF during the recent solar minimum
was also seen in ACE data (Mewaldt 2013).

\begin{figure} [htb]  % Fig 2 scatter plot
   \centering
   \includegraphics[width=11cm]{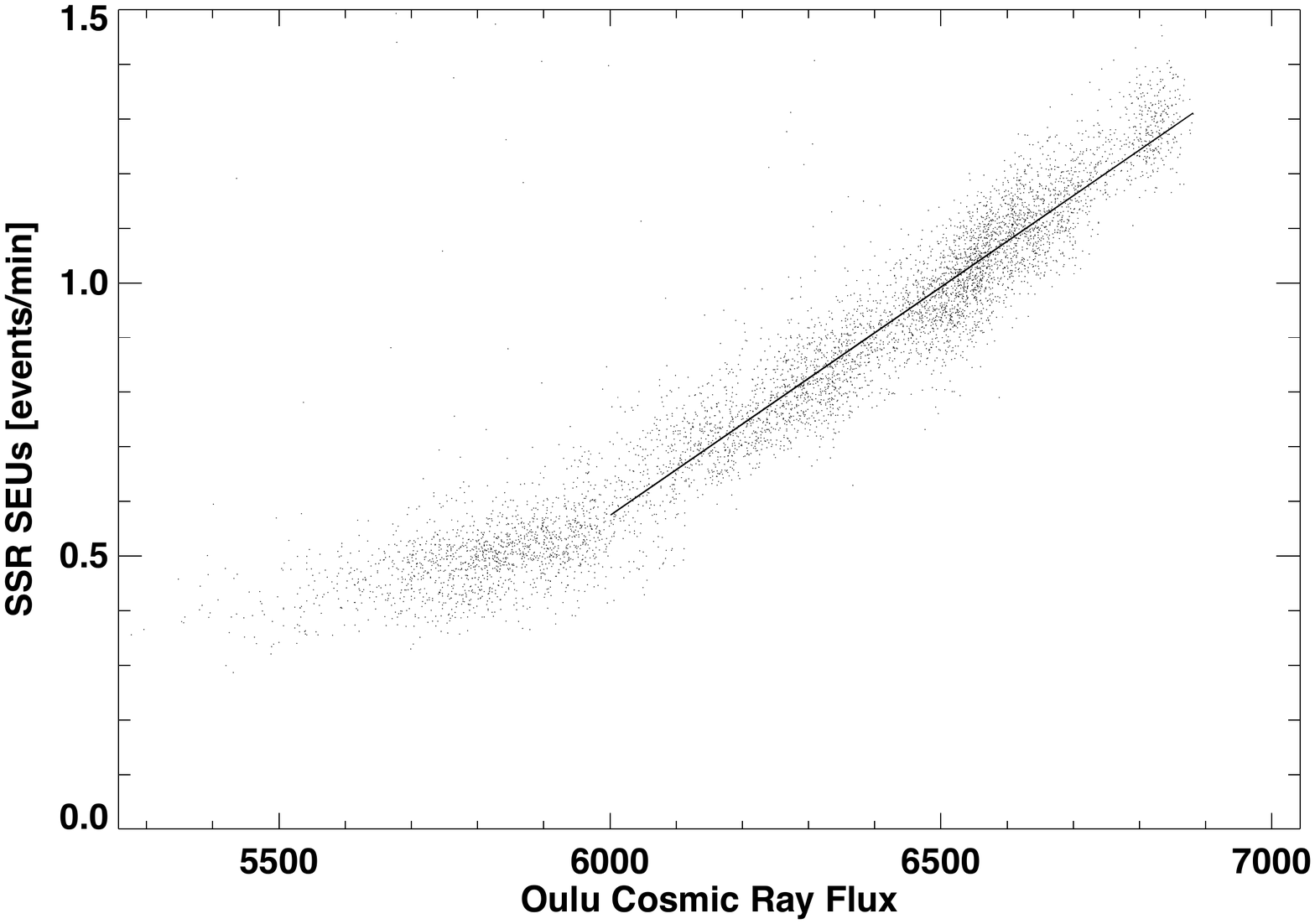}
    \caption{Scatter plot of SOHO SEU events versus CRF seen in Oulu data.
     There is a clear correlation for the population with CRF $>$5900 events.}
\label{scatter}
\end{figure}

To investigate details of the correlation between both data sets, we show a
scatter plot in Fig.~2 for all days without SEP events (daily average SEU rate $<$1.5 cpm).
It is obvious that there are two different populations. The CRF above $\approx$5900
daily events correlates with the SOHO SEU rate above 0.5 cpm. The correlation coefficient
is 0.935 in this linear part. There is almost no correlation in the lower-value section.
About a quarter of the data points fall into this second population.
To demonstrate that these fall into the solar maximum period, we investigate
another detail hardly seen in Fig.~1, namely the spikes in the SEU
data, in particular in the cycle 23 maximum, which are clipped in the plot.

Fig.~3 shows a subset of the same data around solar maximum and demonstrates that the spikes
are only seen in SEU data. These spikes are not of cosmic origin, all of them
coincide with solar storms, e.g. the `Halloween' event on 2003~Oct~31, or the
`Bastille Day' storm on 2000~Jul~14, the strongest one recorded by SOHO
reaching a peak rate of 80~cpm and a daily average of 33~cpm.

\begin{figure} [htb] % Fig 3 zoom
   \centering
   \includegraphics[width=11cm]{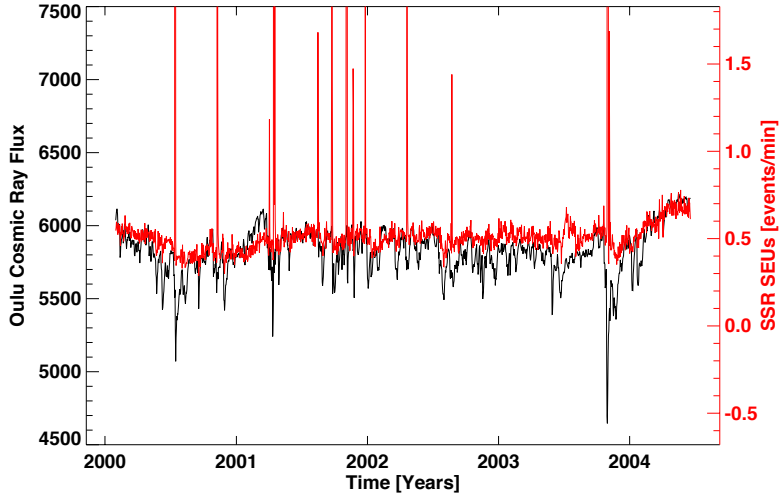}
    \caption{Subset of the data in Fig.~1 during solar maximum.
    The plot shows a dozen sharp spikes on top of the solar-cycle-modulated
    background of SSR SEUs triggered by cosmic ray hits.
    These spikes are caused by isolated strong SEP events. Most of them coincide with
    a CRF down spike.}
\label{zoom}
\end{figure}

Most of the SEU spikes correspond with negative spikes in the
CRF, when the flux drops by up to 25\% over several days. Such events,
known since decades as Forbush decreases, are attributed to magnetic clouds
that fill the inner heliosphere during CME eruptions. Backside CMEs will cause
Forbush decreases without a SEP spike. The SEP spikes are only seen in SSR SEU data,
an effect that can easily be explained by the different energy response of the two
systems. Forbush decreases are only seen in ONM data. This is more difficult
to understand, since SOHO is located in the same magnetic cloud as the
neutron monitors around the globe.

\subsection{Energy response}

Harboe-S{\o}rensen et al. (2002) mention a minimum energy of 10~MeV needed to
cause a SEU in a SSR cell, whereas Bendel and Petersen (1983) report their empirical result
that at least 20--30 MeV protons are required to produce a highly-ionizing track
needed to cause a SEU. We investigated the relationship between the SSR SEU
rate and the solar proton flux at different energies as maesured by GOES.
A good correlation is present for energies above 100~MeV (Fig.~4),
but does not exist for energies $E_{prot} >10$~MeV. This is consistent with
reference information about the shielding of the SSR by 14~mm aluminum.
It requires a proton with an energy of at least 50~MeV to penetrate 14~mm of aluminum
(ESA handbook 2008).

\begin{figure} [htb] % Fig 4 energy
   \centering
   \includegraphics[width=11cm]{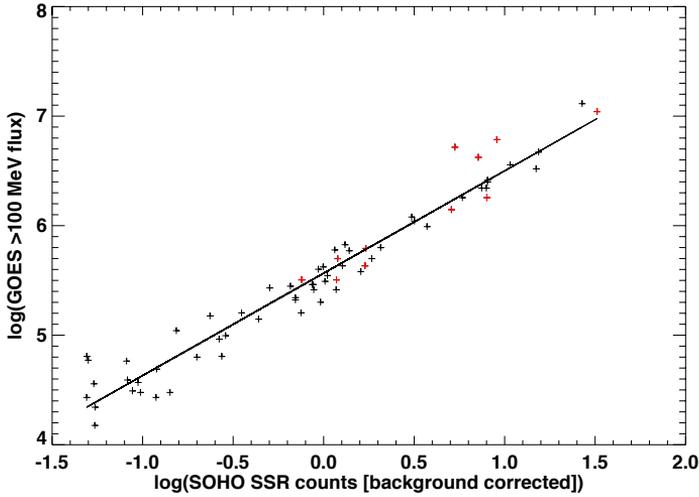}
    \caption{Scatter plot of the GOES100 proton flux ($E_{prot} >100$ MeV)
    versus the SOHO SSR SEUs (CRF background subtracted). The good correlation
    indicates that a proton energy of 100 MeV is sufficient to cause SEUs in the SSR.
    The 11 cases shown in red indicate events that are also seen by neutron monitors.
    Such rare events are called Ground Level Enhancements (GLEs), generated by
    so-called solar cosmic rays in contrast to galactic cosmic rays.
    }
\label{energy}
\end{figure}

We also looked into a scenario where proton-induced nuclear reactions load
the spacecraft with radioactive isotopes. Such isotopes with a half-life time
in the order of one day could easily mimic cosmic ray hits and conceal the
Forbush decrease. A suitable candidate would be the $\beta$-instable isotope Mg-28.
There is, however, no nuclear reaction from Al-27 or Si-28 --- the only abundant species
in the vicinity of the SSR. Therefore, we did not pursue this idea any further.

The SSR data is dominated by particles from the Sun, whereas the ONM is susceptible to
particles with energies $>$500~MeV which are rare in SEPs. But it is these
particles that are reduced in Forbush decreases. Cane et al. (2010) suggested that
SEP events are swamped by low- to mid-energy particles that conceal the Forbush effect.

We added the numbers from all spikes that occurred in the SRR SEU rate over the
entire 17 years. 6\% of the total number are from SEPs and the remaining 94\% are caused by galactic
cosmic rays. During the maximum period of cycle 23 (2000 Jan 1 -- 2003 Dec 31),
the SEP contribution increased to 22\%, and during 2001, the year with the highest SEP rate, to 30\%.

\section{Solar Array Degradation}

Fig.~5 shows the degradation of the solar array efficiency from Dec 1995 until
Feb 2013. The total loss was $\approx$22.5\% during that time (and has reached 24\% at the end of 2014).
The degradation starts with a linear, continuous
decrease of 0.00368\%~/~d (1.344\% per year) from launch to Jul 2000. We attribute
this decrease to the CRF during SOHO's first solar minimum.

\begin{figure*}
   \centering
   \includegraphics[width=14cm]{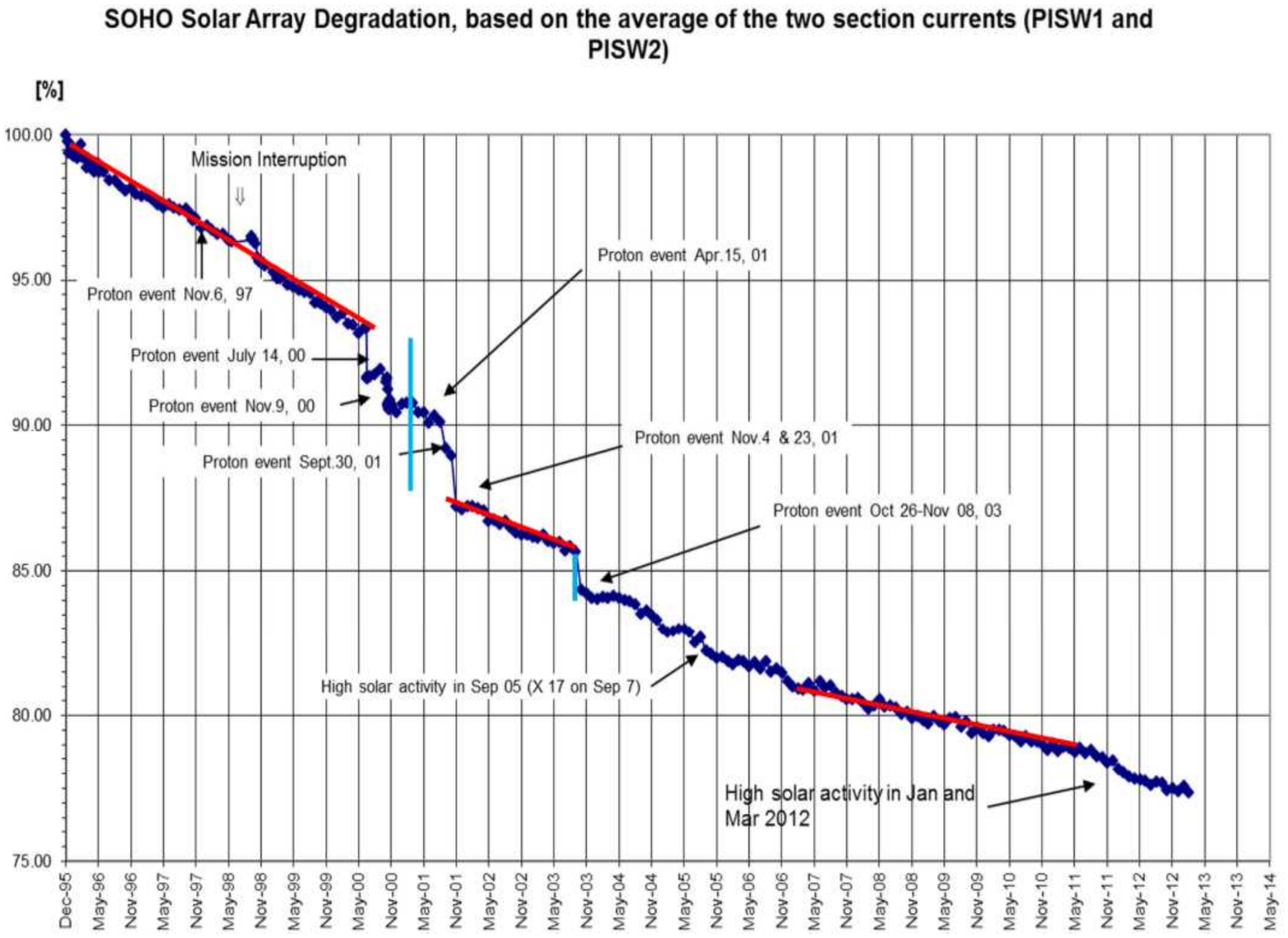}
      \caption{Solar array degradation over the SOHO mission. The plot shows the average of
      both sections in percent of the pre-launch value.
      During early cycle 23 a continuous decrease of 1.34\% per year is observed, which we
      attribute to the CRF during SOHO's first solar minimum.
      There are later episodes with continuous --- but less steep --- decrease.
      Around 2000, individual proton events dominate the scene.
      %Losses of 6\% occurred during 18 months. In total, $\approx$40\% of the degradation
      %must be attributed to proton events. In other words:
      %the effect of a series of violent events is comparable to the integrated effect of the cosmic rays.
      }
\label{Array}
\end{figure*}

Then follows a phase of several stepwise decrements that can
be associated to SEP events during the maximum of cycle 23 around 2001. Here,
individual proton events start to dominate the scene.
Later follow two more episodes with continuous --- but less steep --- decrease. Around
2002, the degradation rate is 0.00284\%~/~d (from a starting point of 87.2\%) and
only 0.00168\%~/~d (from a starting point of 82.1\%) during the period
from Feb 2007 to May 2011. There is no evidence for a significant solar cycle
variation. It seems as if a continuous decrease of the degradation rate
reduces the value by almost a factor of two. We fitted the graph with three linear sections
interrupted by steps and tried to reproduce the
quasi-linear degradation in analogy to the radioactive decay by a e$^{-\lambda t}$ law.
It is obvious that the decay constant, $\lambda$, is not constant during the
observation period. The half-life time, starting from a value of 51.6 years in
1997 increases to 113 years around 2010. We speculate
that in the solar arrays cells of different radiation hardness are found and
that destruction of less-radiation hard cells is in progress all the time.
Also, ageing effects of the cover-glass could be responsible for efficiency
loss.

We tried to quantify the effects of cosmic rays and the effects of SEPs
during this period. In total, of the 22.5\% power loss 8.5\% can be attributed to proton events.
Hereof, 5\% occurred during a period of only 1.5 years. Altogether, 38\% $\pm$ 2\%
of the degradation during 17 years can be attributed to proton events.
In other words:  the effect of a series of violent short-term events on the solar panels
is comparable to the accumulated effect of the CRF over this  period.

\section{Conclusion}

SOHO engineering data provide an excellent record of solar and galactic
cosmic rays that has not yet been exploited for scientific purposes. The SSR SEU
record provides a data archive that exhibits its scientific value in concert with other
particle instruments in space and on the ground.
The main results of this study can be summarized as follows:
\begin{itemize}
\item The SOHO SSR SEU rate and the Oulu cosmic ray flux are highly correlated.
\item SEPs show up as spikes in the SSR SEU data, which allows to discriminate between solar and non-solar effects.
\item  Forbush decreases, which are manifest in the Oulu CRF record, are not visible in the SSR SEU data. This fact is not understood yet.
\item The SOHO SSR is susceptible to particles with energies $\gtrsim$ 100 MeV.
\item The fraction of SSR SEUs caused by SEPs during cycle 23 (1996 Apr 1 -- 2008 Aug 31) is only 6\%. The remaining 94\% are caused by galactic cosmic rays. During solar maximum this fraction increased to 30\%.
\item About 40\% of the total solar array degradation can be attributed to proton events. In other words: the effect of a series of violent events during cycle 23 is comparable to the cycle-integrated damage by cosmic rays.
\end{itemize}

Taking a wider view, it seems that there is no obvious relationship between
particle energy, particle flux, X-ray flux, and CME geoeffectiveness.
The Carrington event was probably the strongest geomagnetic storm in recent history,
but it left no trace in cosmogenic isotopes. The 2005 Jan 20 event was by
far the strongest GLE since we measure cosmic rays, but the related flare and
associated CME was only moderate. This is often confused– not only in public
media, but also by the scientific community.

%\begin{acknowledgements}

\section*{Acknowledgements}
SOHO is a joint mission of ESA and NASA. The support of engineers from the SOHO team
is greatly acknowledged. We specially thank T. van Overbeek and J.-P. Olive
for their contributions to this article.
%\end{acknowledgements}

\section*{References}
\begin{itemize}
\small
\itemsep -3pt
\itemindent -20pt

\item [] Bendel,~W.L. \& Petersen,~E.L.: 1983, {\it ITNS} {\bf 30}, 4481.

\item [] Cane,~H.V., Richardson,~I.G., \& von Rosenvinge,~T.T.: 2010, {\it JGRA}, {\bf 115}, 8101.

\item [] S{\o}rensen,~J. {\it Solar Orbiter Environmental Specification 2.0}:
2008, {\bf TEC-EES-03-034/JS}, 24.

\item [] Harboe-S{\o}rensen,~R., Daly,~E., Teston,~F., Schweitzer,~H., Nartallo,~R., Perol,~P.,
Vandenbussche,~F., Dzitko,~H., \& Cretolle,~J.: 2002, {\it ITNS} {\bf 49}, 1345.

\item [] Mewaldt,~R.A.: 2013, {\it SSRv} {\bf 176}, 365.

\item [] McIntosh,~S.~W., Leamon,~R.~J.,
Gurman,~J.~B., Olive,~J.-P., Cirtain,~J.~W., Hathaway,~D.~H., Burkepile,~J.,
Miesch,~M., Markel,~R.~S., \& Sitongia,~L.: 2013, {\it ApJ} {\bf 765}, 146.

\end{itemize}

\end{document}